\documentclass[conference]{IEEEtran}
\IEEEoverridecommandlockouts
\usepackage{cite}
\usepackage{amsmath,amssymb,amsfonts}
\usepackage{algorithmic}
\usepackage[caption=false, font=footnotesize]{subfig}
\usepackage{graphicx}
\usepackage{textcomp}
\usepackage{url}

\def\BibTeX{{\rm B\kern-.05em{\sc i\kern-.025em b}\kern-.08em
    T\kern-.1667em\lower.7ex\hbox{E}\kern-.125emX}}
\begin{document}

\title{Wideband Channel Estimation for Hybrid Beamforming Millimeter Wave Communication Systems with Low-Resolution ADCs\\
\thanks{This work is supported by gift funding from Huawei Technologies.}
}

\author{\IEEEauthorblockN{Junmo Sung, Jinseok Choi, and Brian L. Evans}
\IEEEauthorblockA{\textit{Wireless Networking and Communications Group} \\
\textit{The University of Texas at Austin}, 
Austin, TX USA \\
junmo.sung@utexas.edu, jinseokchoi89@utexas.edu, bevans@ece.utexas.edu}
}

\maketitle

\begin{abstract}
A potential tremendous spectrum resource makes millimeter wave (mmWave) communications a promising technology. High power consumption due to a large number of antennas and analog-to-digital converters (ADCs) for beamforming to overcome the large propagation losses is problematic in practice. As a hybrid beamforming architecture and low-resolution ADCs are considered to reduce power consumption, estimation of mmWave channels becomes challenging.
We evaluate several channel estimation algorithms for wideband mmWave systems with hybrid beamforming and low-resolution ADCs.
Through simulation, we show that 
1) infinite bit ADCs with least-squares estimation have worse channel estimation performance than do one-bit ADCs with orthogonal matching pursuit (OMP) in an SNR range of interest, 
2) three- and four-bit quantizers can achieve channel estimation performance close to the unquantized case when using OMP, 
3) a receiver with a single RF chain can yield better estimates than that with four RF chains if enough frames are exploited, 
and 4) for one-bit ADCs, exploitation of higher transmit power and more frames for performance enhancement adversely affects estimation performance after a certain point.
\end{abstract}

\section{Introduction} 
\label{sec:introduction}
Millimeter wave (mmWave) multiple-input multiple-output (MIMO) communication is a promising technology for the next generation of cellular networks due to its potential enormous spectrum resource. Thanks to the tiny wavelength of mmWave bands, a large number of antennas can be packed into a small form factor, and at the same time, many antennas are needed in mmWave communication systems to compensate a high propagation loss at such high frequency bands. The systems also need a corresponding number of high-speed analog-to-digital converters (ADCs) which are among the most power-hungry components in radio frequency (RF) processing chains. In order to reduce power consumption by the ADCs, a hybrid digital and analog beamforming architecture is proposed to reduce the number of RF chains while keeping the large number of antennas. Adaptive low-resolution quantizers are also taken into consideration for further power consumption reduction \cite{Choi2017}. 

Using the hybrid beamforming architecture, it is shown that achievable rates of a system equipped with low-resolution quantizers are comparable with that with high-resolution quantizers in the low and medium SNR regimes \cite{Mo2017}. Comparing the hybrid beamforming and digital beamforming architectures with low-resolution ADCs, better spectral vs. energy efficiency trade-off can alternatively be achieved \cite{Abbas2016}. However, channel estimation in such systems becomes more challenging as channels are seen though lenses of analog beamforming and received training signals are distorted due to low-resolution quantizers. 

Channel estimation algorithms for mmWave communication systems can be categorized according to the following criteria: types of beamforming architecture, number of users, and quantization resolution. When it comes to a hybrid beamforming architecture, much of the prior work does not consider low-resolution quantizers. Channel estimation with a single user is addressed in \cite{Venu2017,Rodriguez2017icc,Mendez2015ita,Park2016,Alkh2014} and that with multiple users is in \cite{Gao2016lcom,Hagh2016wsa,Alkh2015icassp}. In the multi-user case, users are equipped with either a single antenna or a single RF chain with analog beamforming. As for all-digital beamforming with low-resolution ADCs, a single user is considered in \cite{Dong2016wcsp,Mo2016,Rusu2015} and multiple users are in \cite{Choi2016}. Only a few papers consider a mmWave hybrid MIMO system with low-resolution quantizers \cite{Rodriguez2016,sung2018iccasp}. References \cite{Venu2017,Rodriguez2017icc,Gao2016lcom,Mo2016} use wideband channels, and the other references \cite{Mendez2015ita,Park2016,Alkh2014,Hagh2016wsa,Alkh2015icassp,Dong2016wcsp,Rusu2015,Choi2016,Rodriguez2016,sung2018iccasp} use narrowband channels.
\begin{figure*}[t]
	\centering
	\includegraphics[width=16cm]{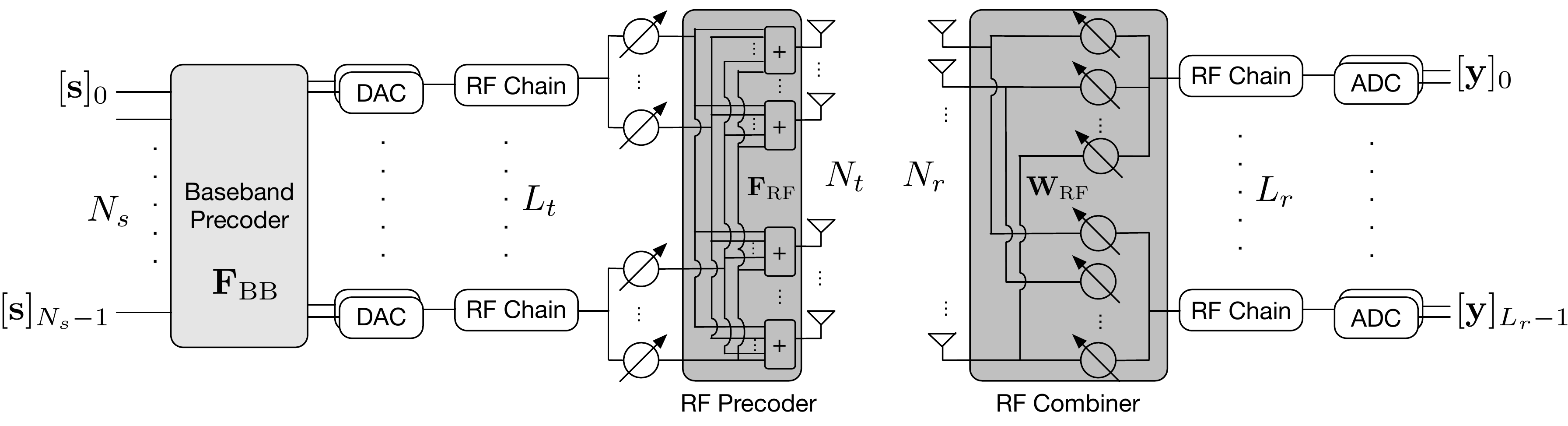}
	\label{fig:system_diagram}
	\caption{Block diagram of a system with the hybrid beamforming architecture and low-resolution ADCs}
\end{figure*}

In this paper, we present the first attempt to estimate wideband channels in a single carrier mmWave MIMO communication system equipped with hybrid digital and analog beamforming and low-resolution ADCs. Such a system is explored in \cite{Rodriguez2016,sung2018iccasp}; however, channels therein are narrowband. The narrowband assumption for mmWave spectrum is not practical as available bandwidths are wide and most likely to be frequency selective. Authors of \cite{Venu2017} consider wideband channels and a hybrid beamforming system model; however, low-resolution quantization is not taken into account. We, therefore, propose to use compressed sensing based algorithms for estimation of wideband channels in a hybrid beamforming and low-resolution mmWave system. 
Simulation results show superior performance of a compressed sensing based algorithm over a traditional channel estimation algorithm; orthogonal matching pursuit (OMP) and least-squares (LS) estimation are considered, respectively, in this paper. In an SNR range of interest, one-bit ADCs with OMP perform better than infinite bit ADCs with LS.
We show that when using OMP, even in a high SNR regime, quantizers with four-bit resolution can achieve estimation performance close to the unquantized case with an NMSE difference less than 2 dB. We also show that exploiting enough frames helps receivers with a single RF chain to yield better channel estimates than those with multiple RF chains. It implicates that small devices that are not capable of accommodating many RF chains can instead use more frames to compensate performance loss. For one-bit quantizers, the results show that higher transmit power and more frames intended for better estimation do not always enhance performance. After a certain point in power and the number of frames, they adversely affect the estimation performance.






\section{System Model} 
\label{sec:system_model}
Consider a wideband mmWave MIMO communication system equipped with a hybrid analog and digital beamforming architecture and low resolution quantizers. A transmitter and a receiver are equipped with $N_t$ and $N_r$ transmit and receive antennas, respectively. The number of RF chains employed at the transmitter and the receiver are $L_t$ and $L_r$, respectively, and $L_t \leq N_t$ and $L_r \leq N_r$. ADCs used at the receiver are assumed to have $b_{\text{ADC}}$-bit resolution, and $b_{\text{ADC}}$ is assumed to not exceed four in this paper. The number of data streams, denoted by $N_s$, can be any integer in $[1, \text{min}(L_t, L_r)]$ and is configured to be the largest for better estimation performance. For simulations represented in this paper, the transmitter and the receiver have the same number of RF chains, i.e., $L_t = L_r$.

We assume that the RF precoder ($\mathbf{F}_{\text{RF}} \in \mathbb{C}^{N_t \times L_t}$) and combiner ($\mathbf{W}_{\text{RF}} \in \mathbb{C}^{N_r \times L_r}$) are implemented with a network of phase shifters. The phase shifters in the system are further assumed to have quantized angles that are represented with $b_{\text{PS}}$ bits. Therefore elements of $\mathbf{F}_{\text{RF}}$ and $\mathbf{W}_{\text{RF}}$ are constrained and expressed as
\begin{align}
	[\mathbf{F}_{\text{RF}}]_{l,m} &= \frac{1}{\sqrt{N_t}} e^{j 2 \pi \frac{k}{2^{b_{\text{PS}}}}}, \\
	[\mathbf{W}_{\text{RF}}]_{l,m} &= \frac{1}{\sqrt{N_r}} e^{j 2 \pi \frac{k}{2^{b_{\text{PS}}}}},
\end{align}
where $[\cdot]_{l,m}$ denotes the element of a matrix in the $l^{th}$ row and the $m^{th}$ column, and $k \in \{0, 1, \cdots, 2^{b_{\text{PS}}}-1\}$. The baseband precoder ($\mathbf{F}_{\text{BB}} \in \mathbb{C}^{L_t \times N_s}$) scales training symbols to maintain the transmit power to be constant. In other words, $\mathbf{F}_{\text{BB}}$ has the constraint such as $\left\lVert \mathbf{F}_{\text{RF}} \mathbf{F}_{\text{BB}} \right\rVert_{F}^2 = N_s$. For simplicity of notation, $\mathbf{F}$ and $\mathbf{W}$ denote $\mathbf{F}_{\text{RF}}\mathbf{F}_{\text{BB}}$ and $\mathbf{W}_{\text{RF}}$, respectively. 

Considering wideband mmWave channels, the channel of the $d^{th}$ delay tap is denoted by $\mathbf{H}_d \in \mathbb{C}^{N_r \times N_t}$ where $d$ is the channel delay and $d = 0, 1, \cdots, N_c - 1$. The received signal in the $m^{th}$ frame can be written as
\begin{align}
	\label{eq:y_first}
	\mathbf{y}_m[n] &= Q\left(\sqrt{\rho} \sum_{d=0}^{N_c-1} \mathbf{W}_{m}^{\mathsf{H}} \mathbf{H}_{d} \mathbf{F}_{m} \mathbf{s}_m[n-d] + \mathbf{W}_{m}^{\mathsf{H}} \mathbf{n}_m[n]\right),
\end{align}
where $Q(\cdot)$ is the quantization operator, $\rho$ is the average received power, $(\cdot)^{\mathsf{H}}$ denotes the conjugate transpose, $\mathbf{s}_m[n] \in \mathbb{C}^{N_{\text{s}} \times 1}$ is the training symbol vector, and $\mathbf{n}_m[n] \in \mathbb{C}^{N_r \times 1} \sim \mathcal{CN}(0, \sigma^2_{\mathbf{n}}\mathbf{I})$ is the noise vector. The training symbol vector satisfies $\mathbb{E}[\mathbf{s}_m[n]\mathbf{s}_m[n]^*] = \frac{1}{N_s}\mathbf{I}$, and symbols are zero if its index is negative. 

For the wideband mmWave channel with multiple paths, the channel at the $d^{th}$ tap can be expressed using a geometric channel model as
\begin{align}
	\label{eq:channel_matrix_full}
	\mathbf{H}_{d} = \sqrt{\frac{N_t N_r}{N_p}}\sum_{l=0}^{N_p-1} \alpha_{l} p(dT_s - \tau_l) \mathbf{a}_{\text{r}}(\theta_{rl}) \mathbf{a}_{\text{t}}^{\mathsf{H}}(\theta_{tl}),
\end{align}
where $\alpha_l \sim \mathcal{CN}(0, \sigma_{\alpha}^2)$ is the complex channel gain of the $l^{th}$ channel path, $p(\tau)$ is the impulse response of the pulse shaping filter evaluated at $\tau$ seconds, $\tau_l$ is the delay of the $l^{th}$ path, $\mathbf{a}_{\text{r}}(\theta_{rl}) \in \mathbb{C}^{N_r \times 1}$ and $\mathbf{a}_{\text{t}}(\theta_{tl}) \in \mathbb{C}^{N_t \times 1}$ are receive and transmit array response vectors associated with the $l^{th}$ path evaluated at $\theta_{rl}$ and $\theta_{tl}$, respectively. $\theta_{rl}$ and $\theta_{tl}$ are azimuth angles of arrival and departure (AoA and AoD) of the $l^{th}$ path and are assumed to be uniform random variables distributed in $[0, 2\pi)$. The variance of the complex channel gain $\sigma_{\alpha}^2$ is such that the channel matrix satisfies $\mathbb{E}[ \left\lVert \mathbf{H}_d \right\lVert_F^2 ] = N_t N_r$. Assuming antennas are in the form of the uniform linear array (ULA), the transmit and receive array response vectors are
\begin{align*}
	\mathbf{a}_{\text{t}}(\theta) &= \sqrt{\frac{1}{N_t}}\left[1, e^{-j2\pi \vartheta}, e^{-j4\pi \vartheta}, \cdots, e^{-j2\pi(N_t-1) \vartheta} \right]^{\mathsf{T}}, \\
	\mathbf{a}_{\text{r}}(\theta) &= \sqrt{\frac{1}{N_r}}\left[1, e^{-j2\pi \vartheta}, e^{-j4\pi \vartheta}, \cdots, e^{-j2\pi(N_r-1) \vartheta} \right]^{\mathsf{T}},
\end{align*}
where $\vartheta=\frac{\mathsf{d}}{\lambda}\sin(\theta)$ is the normalized spatial angle, $\mathsf{d}$ denotes the antenna spacing, $\lambda$ denotes the wavelength, and $(\cdot)^{\mathsf{T}}$ denotes the transpose. The channel matrix in \eqref{eq:channel_matrix_full} can also be written in a more compact form such as
\begin{align}
	\label{eq:compact_form_channel_matrix}
	\mathbf{H}_d = \mathbf{A}_{\text{r}} \mathbf{\Lambda}_{d} \mathbf{A}_{\text{t}}^{\mathsf{H}},
\end{align}
where $\mathbf{\Lambda}_d \in \mathbb{C}^{N_p \times N_p}$ is a square matrix with the scaled complex gains in its diagonal entries, and $\mathbf{A}_{\text{r}} \in \mathbb{C}^{N_r \times N_p}$ and $\mathbf{A}_{\text{t}} \in \mathbb{C}^{N_t \times N_p}$ are matrices with $\mathbf{a}_{\text{r}}(\theta_{rl})$ and $\mathbf{a}_{\text{t}}(\theta_{tl})$ in their columns, respectively. 

\section{Compressed Sensing Channel Estimation} 
\label{sec:compressed_sensing_channel_estimation}
For application of compressed sensing algorithms, we reformulate some of the equations that appear in the previous section. The received signal before quantization in \eqref{eq:y_first} can be rewritten to remove the summation as
\begin{align}
	\mathbf{r}_m[n] = \sqrt{\rho} \mathbf{W}_m^{\mathsf{H}} \mathbf{H} \widetilde{\mathbf{F}}_m \mathbf{s}_m^n + \tilde{\mathbf{n}}_m[n] ,
\end{align}
where $\mathbf{H} = [\mathbf{H}_0 \, \mathbf{H}_1 \, \cdots \, \mathbf{H}_{N_c-1}]$ is the concatenated channel matrix, $\mathbf{s}_m^n = [\mathbf{s}_m[n]^{\mathsf{T}} \, \mathbf{s}_m[n-1]^{\mathsf{T}} \, \cdots \, \mathbf{s}_m[n-N_c+1]^{\mathsf{T}}]^{\mathsf{T}}$, $\widetilde{\mathbf{F}}_m = \mathbf{I}_{N_c} \otimes \mathbf{F}_m$, $\otimes$ denotes the Kronecker product operator, $\mathbf{I}_{N_c}$ is the $N_c \times N_c$ identity matrix, and $\tilde{\mathbf{n}}_m[n]$ denotes $\mathbf{W}_m^{\mathsf{H}} \mathbf{n}_m[n]$. The received frame matrix $\mathbf{R}_m$ whose columns are $\mathbf{r}_m[n]$, $n=0,1,\cdots,N-1$, can then be easily obtained by replacing $\mathbf{s}_m^n$ with $\mathbf{S}_m$ and $\tilde{\mathbf{n}}_m[n]$ with $\mathbf{N}_m$. The matrices, $\mathbf{S}_m$ and $\mathbf{N}_m$, are defined as column-wise concatenated vectors of $\mathbf{s}_m^n$ and $\tilde{\mathbf{n}}_m[n]$, respectively. Namely, 
\begin{align}
	\mathbf{S}_m &= [\mathbf{s}_m^0 \, \mathbf{s}_m^1 \, \cdots \, \mathbf{s}_m^{N-1}], \\
	\mathbf{N}_m &= [\tilde{\mathbf{n}}_m[0] \, \tilde{\mathbf{n}}_m[1] \, \cdots \, \tilde{\mathbf{n}}_m[N-1]].
\end{align}
Stacking $N$ received signal vectors, or equivalently, vectorizing the received frame matrix $\mathbf{R}_m$, we have
\begin{align}
	\label{eq:stacked_rx_vector}
	\mathbf{r}_m &= \left[\mathbf{r}_m[0]^{\mathsf{T}}, \mathbf{r}_m[1]^{\mathsf{T}}, \cdots, \mathbf{r}_m[N-1]^{\mathsf{T}} \right]^{\mathsf{T}}
	= \text{vec}(\mathbf{R}_m) \\
	&= \sqrt{\rho} \left( \mathbf{S}_m^{\mathsf{T}} \widetilde{\mathbf{F}}_m^{\mathsf{T}} \right) \otimes \mathbf{W}^{\mathsf{H}}_m \text{vec} (\mathbf{H})  + \mathbf{v}_m,
\end{align}
where $\mathbf{v}_m = (\mathbf{I}_{N} \otimes \mathbf{W}_m^{\mathsf{H}}) \text{vec} (\mathbf{N}_m)$.
By vectorizing the channel matrix $\mathbf{H}_d$ in \eqref{eq:compact_form_channel_matrix}, we obtain
\begin{align}
	\text{vec}(\mathbf{H}_d) &= (\mathbf{A}_{\text{t}}^* \circ \mathbf{A}_{\text{r}}) 
	\begin{bmatrix}
		[\mathbf{\Lambda}_d]_{0,0} \\
		[\mathbf{\Lambda}_d]_{1,1} \\
		\vdots \\
		[\mathbf{\Lambda}_d]_{N_p,N_p} \\
	\end{bmatrix}  \\
	&= (\mathbf{A}_{\text{t}}^* \circ \mathbf{A}_{\text{r}}) \text{diag}(\mathbf{\Lambda}_d),
	\label{eq:vectorized_channel_matrix}
\end{align}
where $\circ$ denotes the Khatri-Rao product operator, $(\cdot)^{*}$ denotes the complex conjugation, and $\text{diag}(\cdot)$ denotes the operator that extracts diagonal elements of a square matrix and create a column vector. Using \eqref{eq:vectorized_channel_matrix}, vectorization of concatenated channel matrix $\mathbf{H}$ is obtained as 
\begin{align}
	\label{eq:vec_concat_channel_1}
 	\text{vec}(\mathbf{H}) = \left(\mathbf{I}_{N_c} \otimes \mathbf{A}_{\text{t}}^{*} \circ  \mathbf{A}_{\text{r}} \right) 
 	\begin{bmatrix}
 		\text{diag} \left( \mathbf{\Lambda}_0\right) \\
 	    \text{diag} \left( \mathbf{\Lambda}_1\right) \\
 	    \cdots \\
 	    \text{diag} \left( \mathbf{\Lambda}_{N_c-1}\right) \\
 	\end{bmatrix}.
\end{align}
\begin{figure}[t]
	\centering
	\includegraphics[width=7cm]{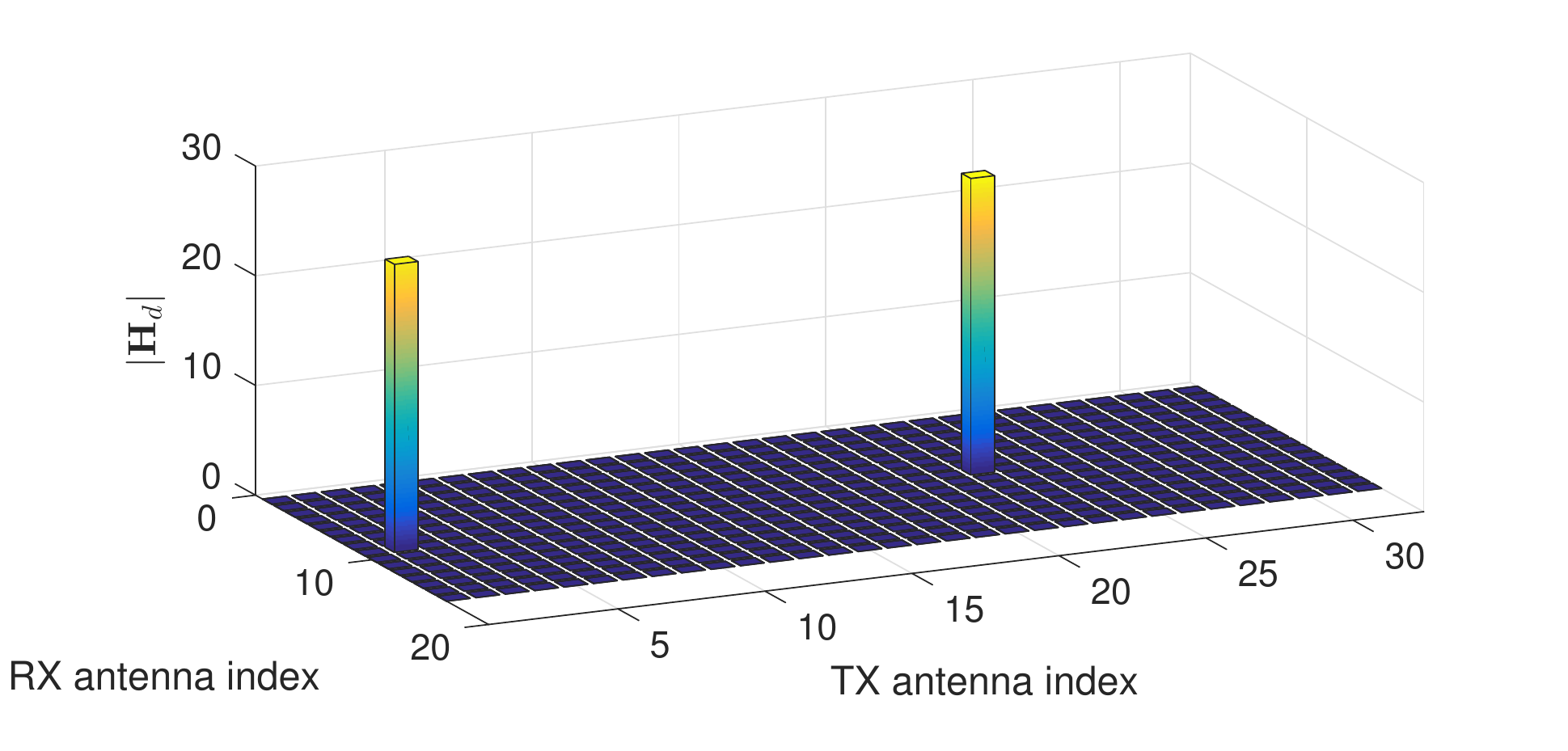}
	\caption{Magnitude of virtual channel matrix without leakage effect. Bins that do not contain channel paths have zero magnitude.}
	\label{fig:virtual_channel_plot_no_leakage}
\end{figure}
\begin{figure}[t]
	\centering
	\subfloat[]{
		\label{fig:virtual_channel_plot_with_leakage_1}
		\includegraphics[width=7cm]{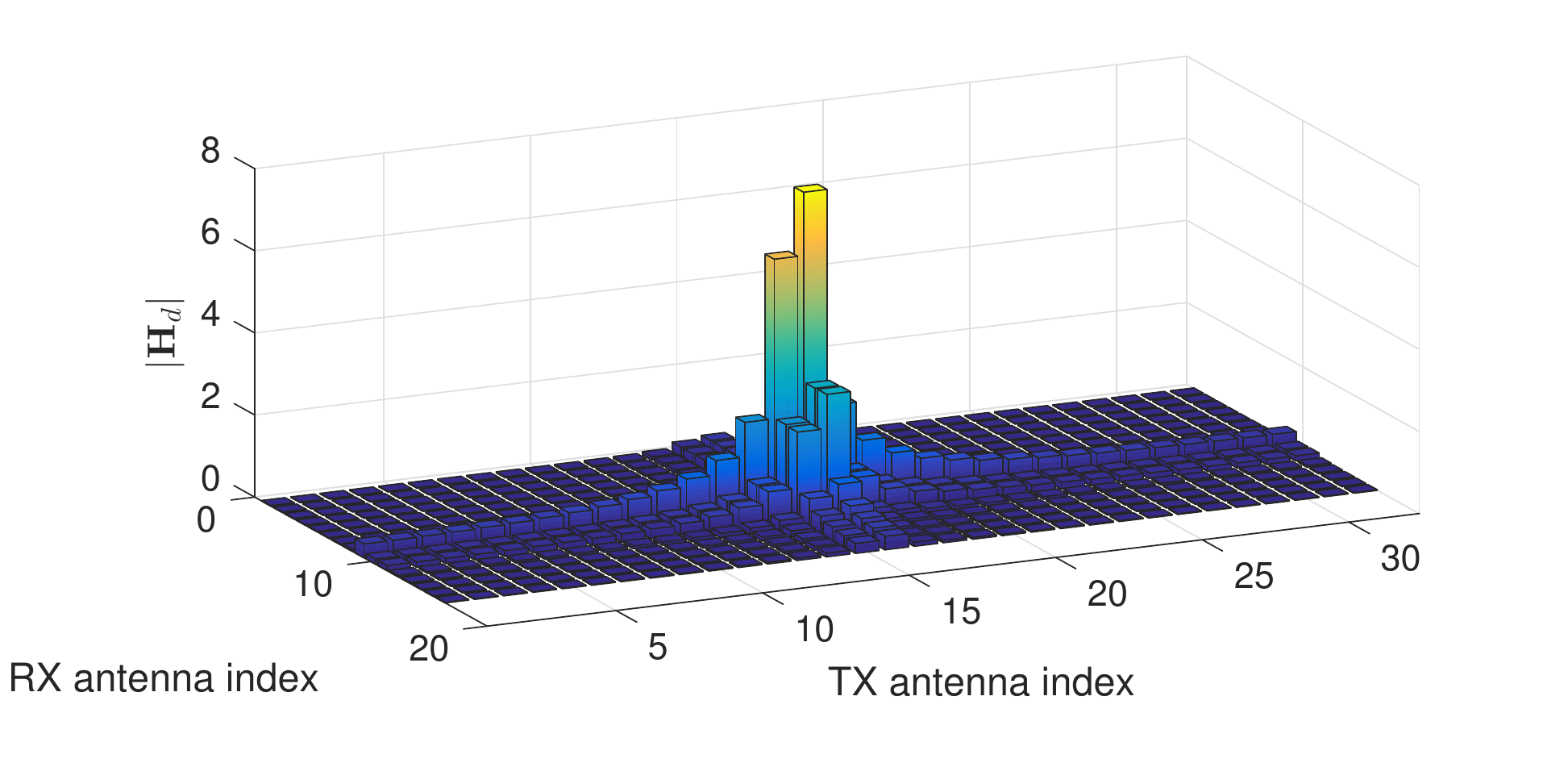}} \\
	\subfloat[]{
		\label{fig:virtual_channel_plot_with_leakage_2}
		\includegraphics[width=7cm]{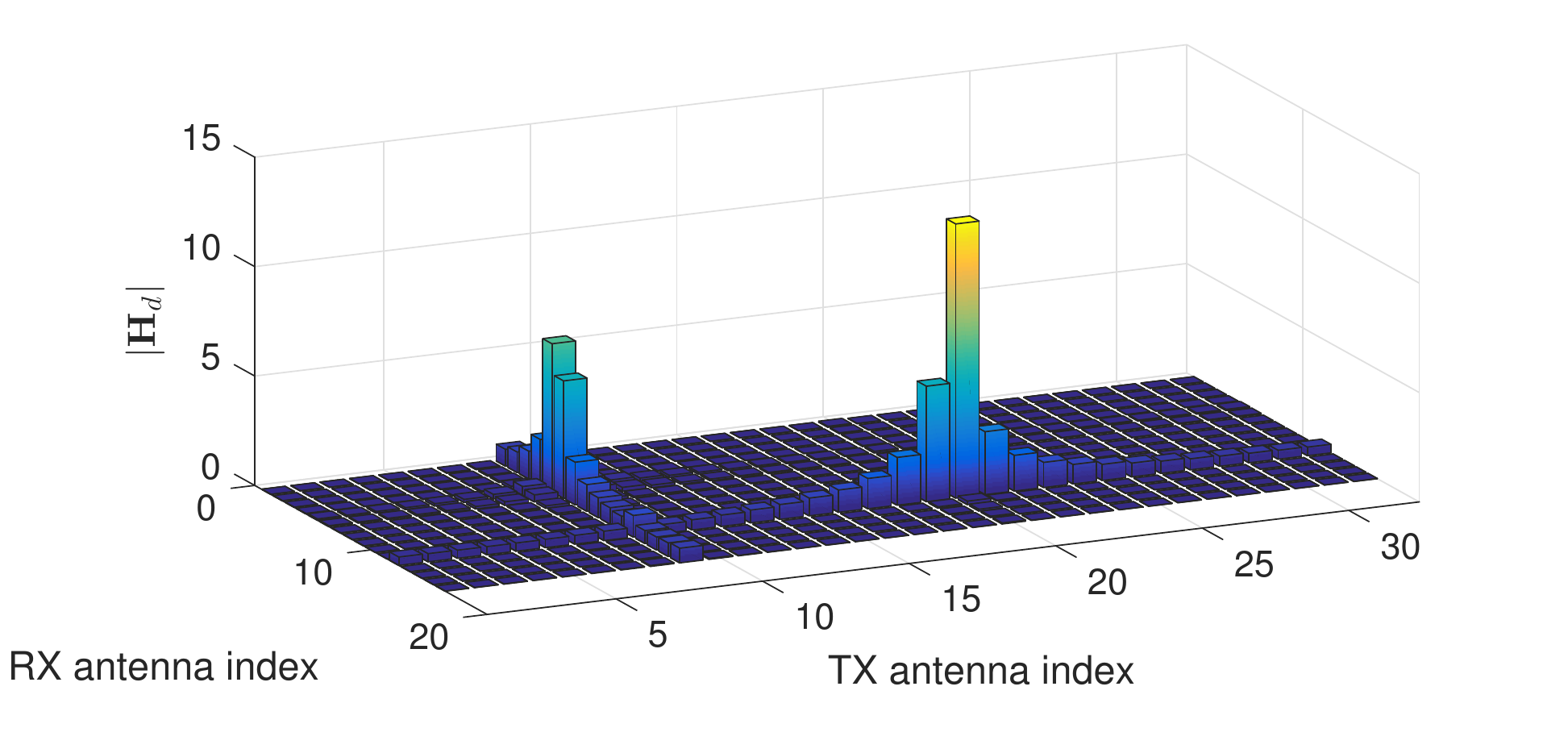}}
	\caption{Magnitude of virtual channel matrix with leakage effect. Both (a) and (b) show a channel with two paths. In (a), one of the paths is smaller and hidden in spreads of the other. In (b), two paths are clearly separated; however, one of the paths might have a smaller magnitude than spreads of the other path.}
	\label{fig:virtual_channel_with_leakage}
\end{figure}

As AoA's and AoD's are unknown, using the virtual channel representation, \eqref{eq:vec_concat_channel_1} can also be represented as
\begin{align}
	\label{eq:channel_vector_virtual_channel}
    \text{vec}(\mathbf{H}) = \left(\mathbf{I}_{N_c} \otimes \mathbf{U}_{\text{t}}^{*} \otimes \mathbf{U}_{\text{r}} \right) \mathbf{h},
\end{align}
where $\mathbf{U}_{\text{t}}$ and $\mathbf{U}_{\text{r}}$ are the truncated Discrete Fourier Transform (DFT) matrices with a size of $N_t \times G_t$ and $N_r \times G_r$, respectively, and $\mathbf{h} \in \mathbb{C}^{N_c G_t G_r \times 1}$ is a sparse vector that contains diagonal elements of $\mathbf{\Lambda}_d$'s. In other words, $\mathbf{h}$ is equivalent to the vectorization of the concatenated virtual channel matrices. With the help of the Kronecker product and the DFT matrices that are larger than the array response matrices, $\mathbf{h}$ can become sparse. It should be noted that sparsity of $\mathbf{h}$ does not necessarily coincide with $N_p$ since actual AoA's and AoD's on which columns of $\mathbf{A}_{\text{t}}$ and $\mathbf{A}_{\text{r}}$ are based do not have to be along with the grids of angles generated by DFT matrices. Virtual channels seen through lenses of an RF precoder and a combiner are, thus, highly likely to have leakage spreading to adjacent bins. Fig.~\ref{fig:virtual_channel_plot_no_leakage} shows the magnitude of a virtual channel at a given moment with two channel paths without such leakage effect. The two paths located in two bins are clearly seen, and adjacent bins have zero magnitude. Fig.~\ref{fig:virtual_channel_with_leakage} illustrates two different cases that can happen to virtual channels with two channel paths due to the leakage effect. In Fig.~\ref{fig:virtual_channel_plot_with_leakage_1}, one of the paths is hidden in a spread of a stronger path. Even if a receiver is assumed to know the number of existing channel paths -- it is two in Fig.~\ref{fig:virtual_channel_plot_with_leakage_1} -- the smaller magnitude path cannot be distinguished from the stronger path that has spatial beam leakage. In contrast, Fig.~\ref{fig:virtual_channel_plot_with_leakage_2} shows two paths that are clearly separated. The magnitude of the weaker path, however, is comparable with spreads of the stronger path. It possibly results in erroneous detection of the weaker path. 

By plugging \eqref{eq:channel_vector_virtual_channel} into \eqref{eq:stacked_rx_vector}, the vectorized received frame is rewritten as
\begin{align}
	\mathbf{r}_m = \sqrt{\rho} \left(\mathbf{S}_m^{\mathsf{T}}  \widetilde{\mathbf{F}}_m^{\mathsf{T}} \right) \otimes \mathbf{W}^{\mathsf{H}}_m  \left( 	\mathbf{I}_{N_c} \otimes \mathbf{U}_{\text{t}}^{*} \otimes \mathbf{U}_{\text{r}} \right) \mathbf{h}  + \mathbf{v}_m.
\end{align}
The received signals of multiple frames can also be stacked and expressed as
\begin{align}
	\mathbf{r} = \left[\mathbf{r}_0^{\mathsf{T}}, \mathbf{r}_1^{\mathsf{T}}, \cdots, \mathbf{r}_{M-1}^{\mathsf{T}} \right]^{\mathsf{T}}
	= \sqrt{\rho} \mathbf{\Phi} \mathbf{\Psi} \mathbf{h} + \mathbf{v},
\end{align}
where $\mathbf{\Phi}$, $\mathbf{\Psi}$ and $\mathbf{v}$ are defined as
\begin{align}
	\mathbf{\Phi} = 
	\begin{bmatrix}
		\mathbf{S}_0^{\mathsf{T}} \widetilde{\mathbf{F}}_0^{\mathsf{T}} \otimes \mathbf{W}_0^{\mathsf{H}} \\
		\mathbf{S}_1^{\mathsf{T}} \widetilde{\mathbf{F}}_1^{\mathsf{T}} \otimes \mathbf{W}_1^{\mathsf{H}} \\
		\vdots \\
		\mathbf{S}_{M-1}^{\mathsf{T}} \widetilde{\mathbf{F}}_{M-1}^{\mathsf{T}} \otimes \mathbf{W}_{M-1}^{\mathsf{H}} \\
	\end{bmatrix},
\end{align}
\begin{align*}
	\mathbf{\Psi} = \mathbf{I}_{N_c} \otimes \mathbf{U}_{\text{t}}^{*} \otimes \mathbf{U}_{\text{r}} , \text{ and }
	\mathbf{v} = \left[\mathbf{v}_0^{\mathsf{T}}, \mathbf{v}_1^{\mathsf{T}}, \cdots, \mathbf{v}_{M-1}^{\mathsf{T}} \right]^{\mathsf{T}}.
\end{align*}
Finally, the quantized received signal $\mathbf{y}$ can be written as
\begin{align}
	\mathbf{y} = Q \left( \mathbf{r} \right) = Q \left( \sqrt{\rho} \mathbf{\Phi} \mathbf{\Psi} \mathbf{h} + \mathbf{v} \right),
\end{align}
where the uniform mid-rise ADCs with $b_{ADC}$-bit resolution are used for quantization. 
\begin{figure}[t]
	\centering
	\includegraphics[width=8.5cm]{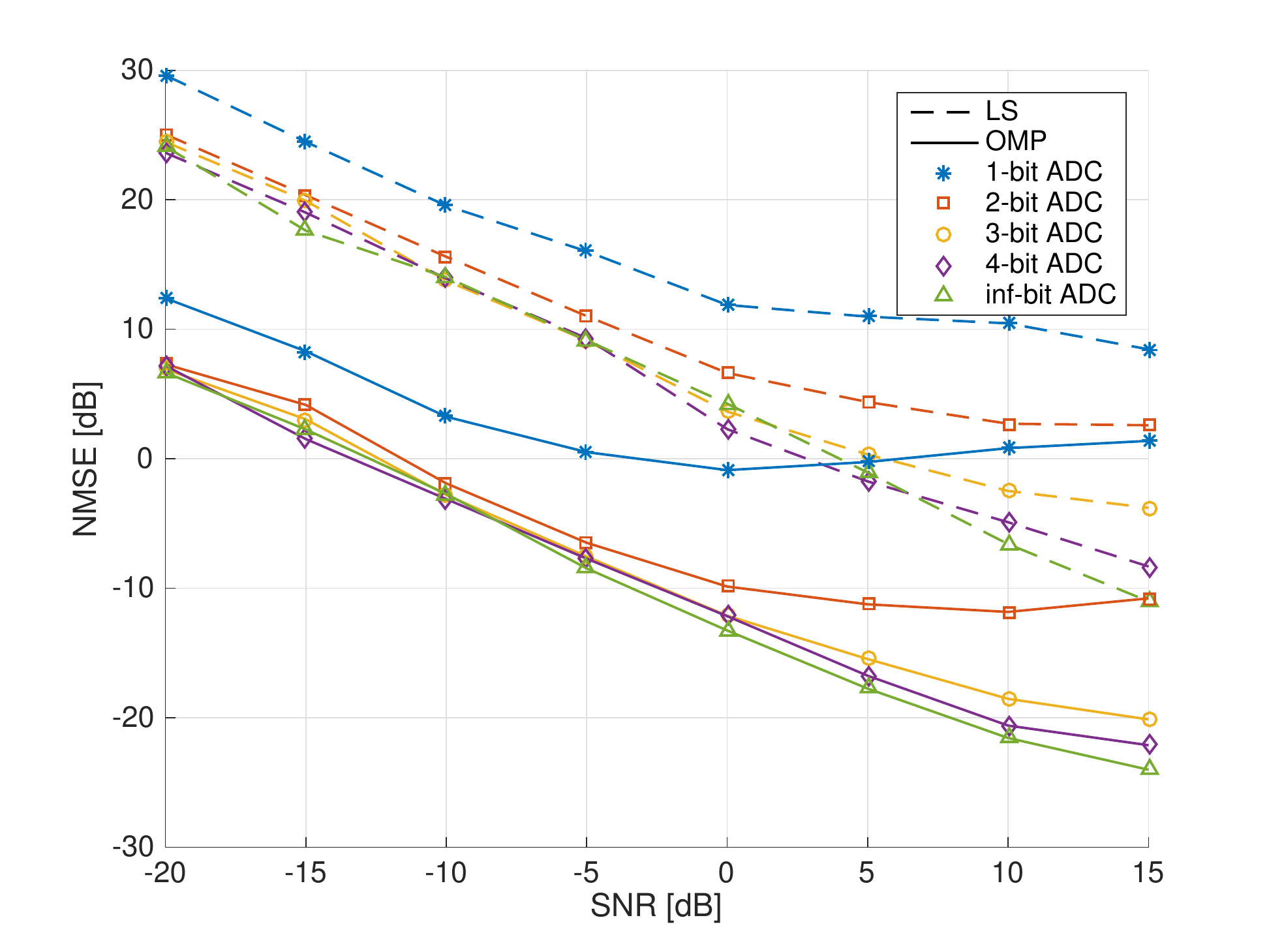}
	\caption{Channel estimation error as a function of SNR. Resolution of 1--4 and infinite bits and channel estimation algorithms of LS and OMP are plotted. Four RF chains and 80 frames are used for estimation.}
	\label{fig:ls_vs_omp}
\end{figure}

\section{Numerical Results} 
\label{sec:numerical_results}
With reformulation of the received signal for sparse recovery, we consider OMP as a tool to recover the sparse channel vector from the received signals. Generalized Approximate Message Passing (GAMP) and its variants are considered for channel estimation in mmWave communication systems \cite{Mo2014,Mo2016}; however, they do not work well in this case especially when many antennas are considered. The reason is that GAMP is not guaranteed to converge for measurement matrices whose peak-to-average ratio of the squared singular values is high \cite{rangan2014isit}. The unit-magnitude constraints on RF beamforming matrices make a measurement matrix ill-conditioned.  
In this section, channel estimation performance of OMP is evaluated in several aspects and is compared with that of LS estimation. The Matlab code is available at \cite{projectcode}. The performance metric is the NMSE which is defined as
\begin{align}
	\text{NMSE} = \mathbb{E}\left[ \frac{\lVert \mathbf{H} - \hat{\mathbf{H}}\rVert_{F}^2}{\lVert \mathbf{H} \rVert_{F}^2} \right],
\end{align}
where $\hat{\mathbf{H}}$ is the recovered channel matrix using the channel estimation algorithms. Parameters used in this section are as follows unless otherwise stated for a specific figure: $N_t=32$, $G_t=64$ $N_r=16$, $G_r=32$, $L_t=L_r=N_s=4$, $N_c=4$, $N_p=2$, $M=80$, $N=16$ and $b_{\text{PS}}=6$. 
\begin{figure}[t]
	\centering
	\subfloat[]{
		\label{fig:nmse_Infbit_M_10_100_Lt_1_2_4}
		\includegraphics[width=8.5cm]{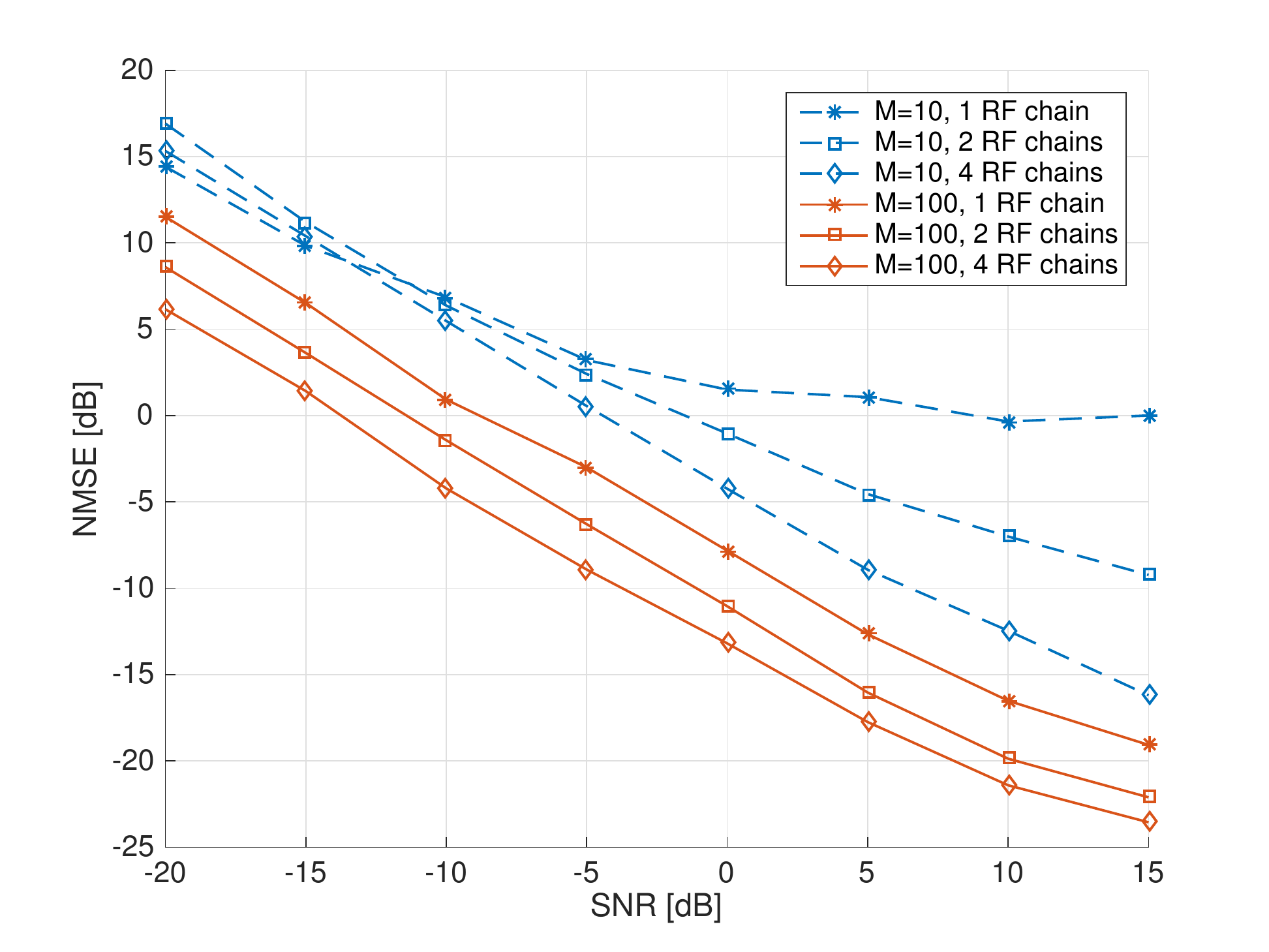}} \\
	\subfloat[]{
		\label{fig:nmse_4bit_M_10_100_Lt_1_2_4}
		\includegraphics[width=8.5cm]{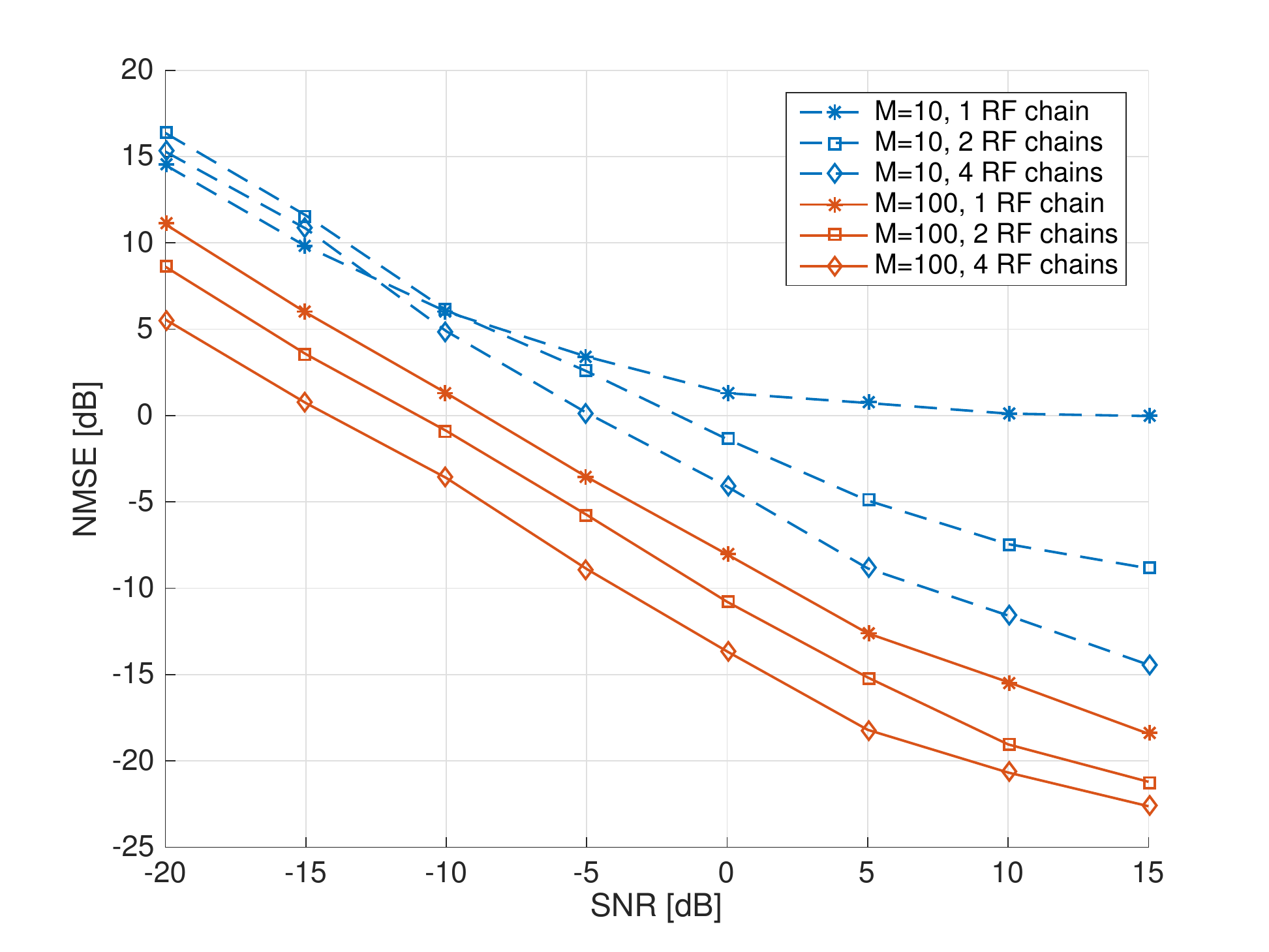}}
	\caption{Channel estimation error for combinations of $M=\{10, 100\}$ and $L_t=L_r=\{1, 2, 4\}$. Infinite bit and four-bit ADCs are used in (a) and (b), respectively.}
	\label{fig:nmse_var_snr_M_Lr}
\end{figure}

Fig.~\ref{fig:ls_vs_omp} plots NMSE in dB scale using LS and OMP. Infinite bits in this section refers to the resolution of ADCs that does not perform quantization on its input. For both LS and OMP, performance gaps between infinite and low resolution become greater as SNR increases, and finer resolution performs better than coarser one. Comparing LS and OMP, OMP performs better than LS for each resolution considered in the figure. Across an SNR range from -20 to 15 dB, LS with infinite bit resolution has worse performance than does OMP with two bits. Considering the fact that SNR is generally low during a channel estimation stage and assuming SNR is lower than 5 dB, one-bit ADCs with OMP yield better estimates than infinite bit with LS. 

Focusing on OMP, NMSE of three- or four-bit resolution is close to that of infinite bit resolution. For four-bit ADCs, NMSE difference is still less than 2 dB even at an SNR of 15 dB. For two-bit ADCs, channel estimation performance is comparable with that with higher resolution ADCs in a low SNR regime (less than 0 dB in Fig.~\ref{fig:ls_vs_omp}) and is far better than that with infinite bit ADCs using LS.
It is worth mentioning that, for one-bit ADCs with OMP, NMSE decreases with the increase in SNR up until a certain point, and then it reversely increases. This phenomenon is known as stochastic resonance \cite{mcdonnell2008stochastic,Mo2014}. 

\begin{figure}[t]
	\centering
	\includegraphics[width=8.5cm]{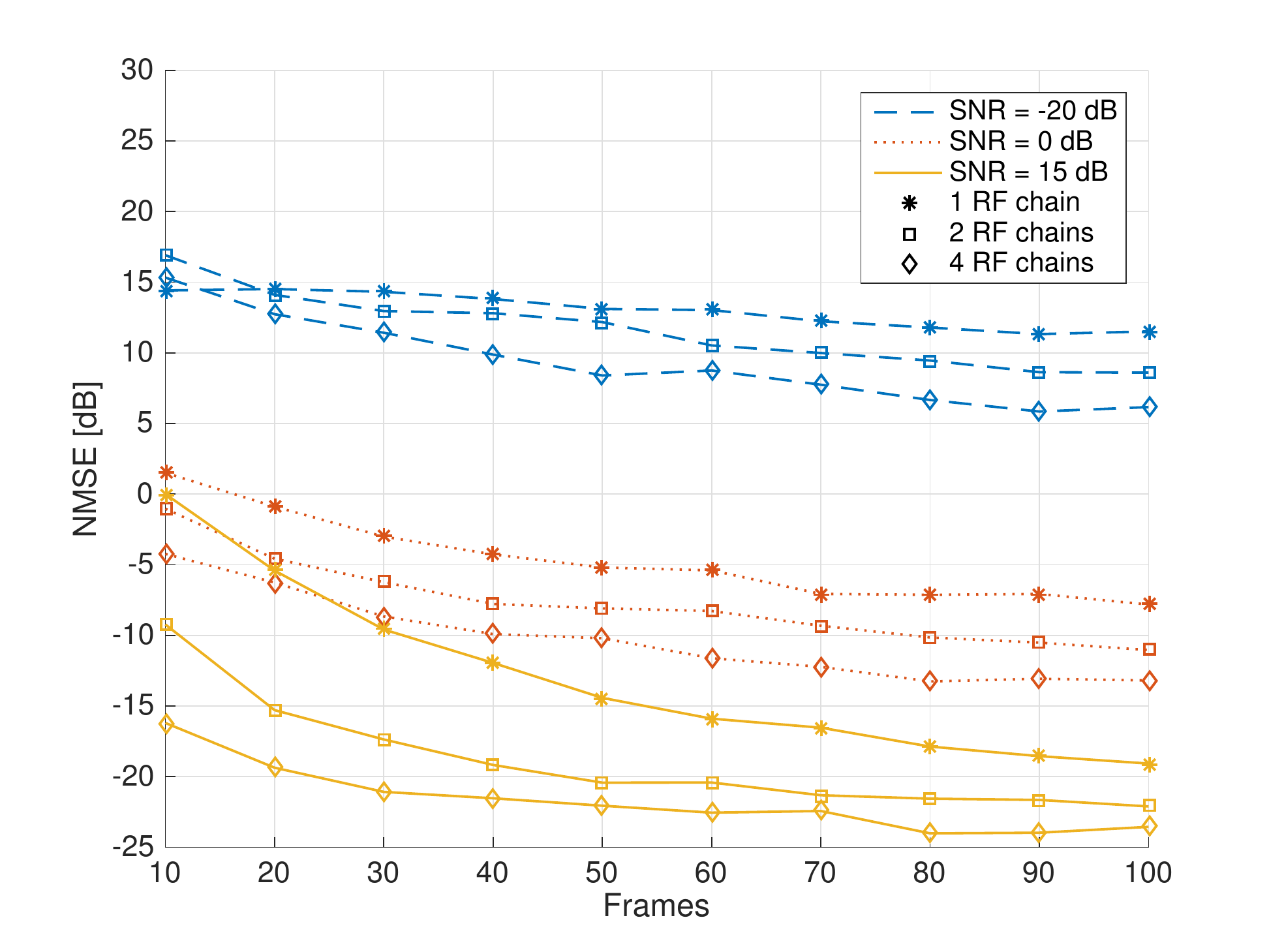}
	\caption{Channel estimation error as a function of the number of frames. Combinations of SNR=$\{-20, 0, 15\}$ [dB] and $L_t=L_r=\{1, 2, 4\}$ are plotted. Quantization resolution is infinite, and 80 frames are used.}
	\label{fig:nmse_var_frame_inf_bits}
\end{figure}
\begin{figure}[t]
	\centering
	\includegraphics[width=8.5cm]{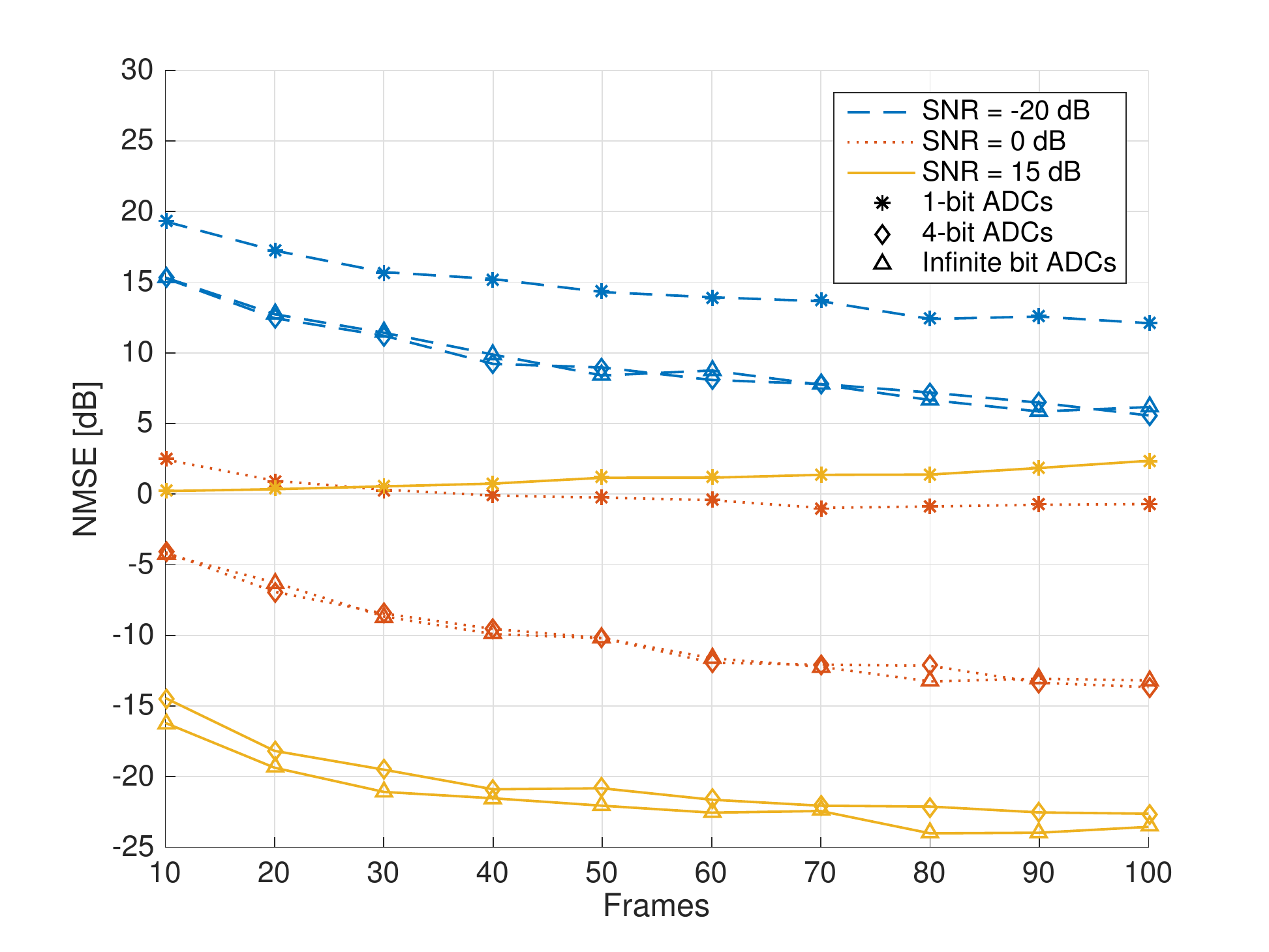}
	\caption{Channel estimation error as a function of the number of frames. Combinations of SNR=$\{-20, 0, 15\}$ [dB] and $b_{ADC}=\{1, 4, \infty\}$ are plotted. The number of RF chains is fixed at four, and 80 frames are used.}
	\label{fig:nmse_var_frame_Lt_4}
\end{figure}
Fig.~\ref{fig:nmse_var_snr_M_Lr} is provided to show differences in channel estimation performance caused by the number of RF chains and the number of frames. In both Figs.~\ref{fig:nmse_Infbit_M_10_100_Lt_1_2_4} and \ref{fig:nmse_4bit_M_10_100_Lt_1_2_4}, combinations of $M=\{10, 100\}$ and $L_t=L_r=\{1, 2, 4\}$ are plotted while Fig.~\ref{fig:nmse_Infbit_M_10_100_Lt_1_2_4} is for infinite bit resolution and Fig.~\ref{fig:nmse_4bit_M_10_100_Lt_1_2_4} is for four-bit resolution. Comparing two subfigures, it is first noted that performance gaps between four-bit and infinite bit resolution are small as shown earlier in Fig.~\ref{fig:ls_vs_omp}. Channel estimation using a hundred frames performs better than that with ten frames. Effect of the number of frames on estimation performance is elaborated more in detail in the following paragraph. It can also be seen that more RF chains generally achieve lower estimation error. Considering transceivers equipped with a single RF chain, if the receiver uses enough frames for estimation, performance can be enhanced and becomes better than that with four RF chains as shown in both subfigures.

Effect of the number of frames on channel estimation error is illustrated in Figs.~\ref{fig:nmse_var_frame_inf_bits} and \ref{fig:nmse_var_frame_Lt_4}. Assuming that quantizers have infinite bit resolution, Fig.~\ref{fig:nmse_var_frame_inf_bits} shows estimation errors as a function of the number of frames. Several SNR values and numbers of RF chains are plotted. When quantization is perfect, the figure shows, in general, more frames, more RF chains, and higher SNR are beneficial to channel estimation. This statement is not always true if quantization is not perfect. Fig.~\ref{fig:nmse_var_frame_Lt_4} shows estimation errors as with Fig.~\ref{fig:nmse_var_frame_inf_bits} whereas the number of RF chains is fixed at four and various quantization resolution is considered. It can be seen that exploitation of more frames helps OMP to achieve lower estimation error with an exception: one-bit resolution at high SNR. One-bit quantization curves in Fig.~\ref{fig:nmse_var_frame_Lt_4} for SNR of 0 and 15 dB indicate that more frames and higher SNR do not necessarily guarantee better estimation performance unlike the perfect quantization case. When using more than 30 frames, estimation error for an SNR of 15 dB is greater than that for an SNR of 0 dB. For an SNR of 15 dB, using more frames adversely affects estimation performance. This phenomenon is similar to the stochastic resonance shown in Fig.~\ref{fig:ls_vs_omp}. With four-bit resolution, the estimation performance is close to that with infinite resolution across all SNR and frame ranges shown in the figure, and the performance gap between the two resolution expands with higher SNR. 

\begin{figure}[t]
	\centering
	\includegraphics[width=8.5cm]{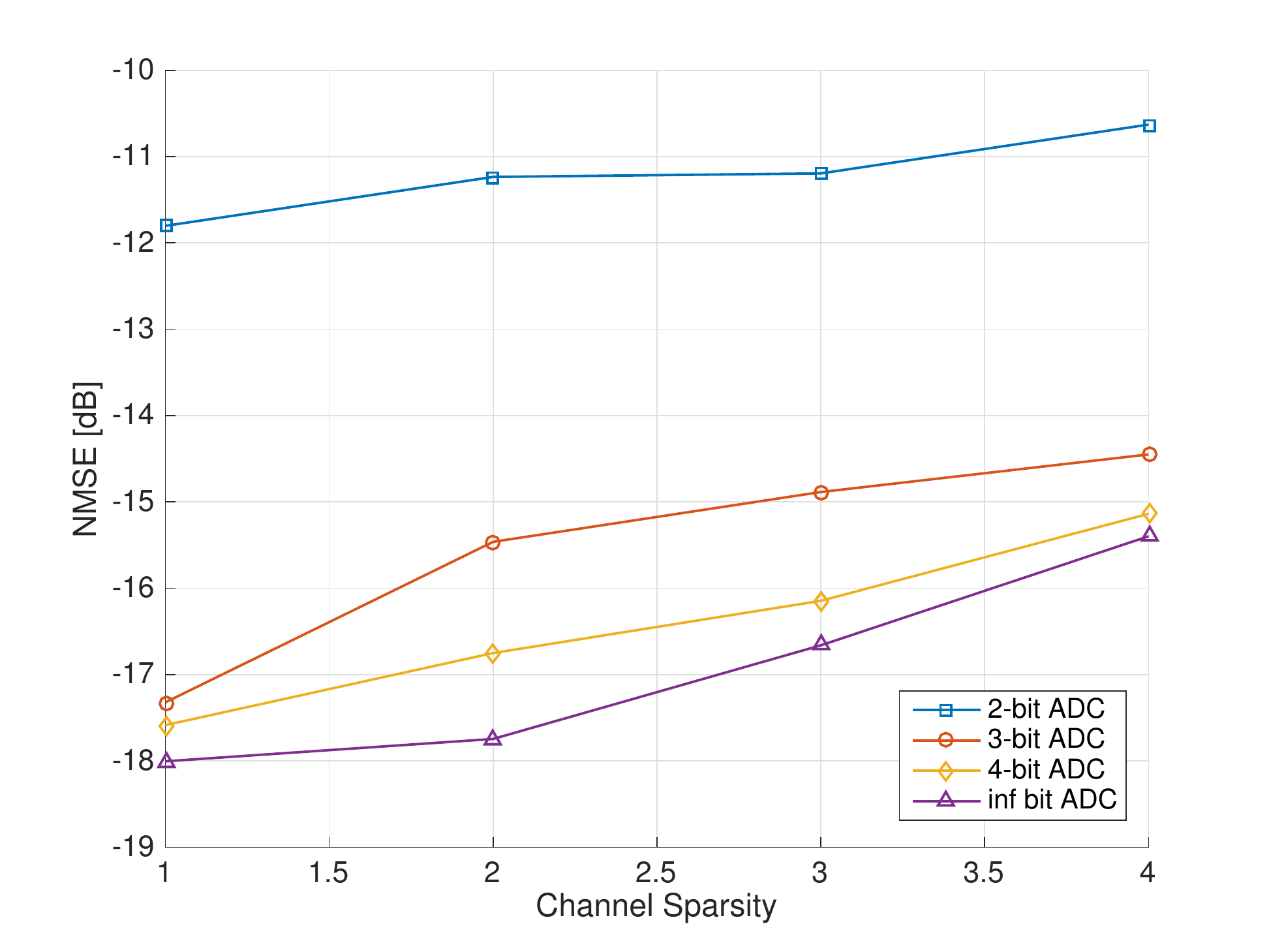}
	\caption{Channel estimation error as a function of channel sparsity. Two through four and infinite bit resolution at an SNR of 5 dB is plotted. Four RF chains and 80 frames are considered. }
	\label{fig:nmse_var_sparsity}
\end{figure}

Fig.~\ref{fig:nmse_var_sparsity} shows NMSE simulated with four RF chains and 80 frames. Resolution of two through four and infinite bits is plotted. The sparsity in this figure is equivalent to the number of channel paths $(N_p)$. As the sparsity increases, channel estimation performance degrades for all quantization resolution under consideration. Comparing four curves in the figure, it is seen that finer resolution decreases estimation error. 

\section{Conclusion} 
\label{sec:conclusion}
In this paper, we presented the first attempt to estimate wideband channels using a compressed sensing based algorithm in a mmWave communication system equipped with hybrid beamforming and low-resolution quantizers. 
We compared performance between OMP and LS as channel estimation algorithms and showed that one-bit quantizers using OMP perform better than infinite bit quantizers using LS in an SNR rage of interest. The results showed that, when using OMP, estimation performance of infinite bit quantizers can closely be achieved with four-bit quantizers even in a high SNR regime with less than a 2 dB estimation error gap in terms of NMSE. Across all ranges of SNR, RF chains, frames and sparsity considered in this paper, four-bit ADCs were shown to achieve performance on par with infinite bit ADCs. 
Performance loss due to the limited number of RF chains on small devices can be compensated by exploiting more frames. It was reported that, after a certain point, using higher transmit power adversely affects the channel estimation performance of a system with one-bit quantizers \cite{Mo2014,Stockle2016}. We showed that this is also the case when using more frames at high SNR.

\bibliographystyle{IEEEbib}
\bibliography{ICC_OMP}

\begin{thebibliography}{10}

\bibitem{Choi2017}
J.~Choi, B.~L. Evans, and A.~Gatherer,
\newblock ``Resolution-adaptive hybrid {MIMO} architectures for millimeter wave
  communications,''
\newblock {\em IEEE Trans. Signal Process.}, vol. 65, no. 23, pp. 6201--6216,
  Dec. 2017.

\bibitem{Mo2017}
J.~Mo, A.~Alkhateeb, S.~Abu-Surra, and R.~W. Heath,
\newblock ``Hybrid architectures with few-bit {ADC} receivers: Achievable rates
  and energy-rate tradeoffs,''
\newblock {\em IEEE Trans. Wireless Commun.}, vol. 16, no. 4, pp. 2274--2287,
  Apr. 2017.

\bibitem{Abbas2016}
W.~B. Abbas, F.~Gomez-Cuba, and M.~Zorzi,
\newblock ``Millimeter wave receiver efficiency: A comprehensive comparison of
  beamforming schemes with low resolution {ADCs},''
\newblock {\em IEEE Trans. Wireless Commun.}, to be published.

\bibitem{Venu2017}
K.~Venugopal, A.~Alkhateeb, N.~Gonz{\'{a}}lez-Prelcic, and R.~W. Heath,
\newblock ``Channel estimation for hybrid architecture-based wideband
  millimeter wave systems,''
\newblock {\em IEEE J. Sel. Areas in Commun.}, vol. 35, no. 9, pp. 1996--2009,
  Sept. 2017.

\bibitem{Rodriguez2017icc}
J.~Rodr{\'{i}}guez-Fern{\'{a}}ndez, K.~Venugopal, N.~Gonz{\'{a}}lez-Prelcic,
  and R.~W. Heath,
\newblock ``A frequency-domain approach to wideband channel estimation in
  millimeter wave systems,''
\newblock in {\em Proc. IEEE Int. Conf. on Commun.}, May 2017, pp. 1--7.

\bibitem{Mendez2015ita}
R.~M{\'{e}}ndez-Rial, C.~Rusu, A.~Alkhateeb, N.~Gonz{\'{a}}lez-Prelcic, and
  R.~W. Heath,
\newblock ``Channel estimation and hybrid combining for {mmWave}: Phase
  shifters or switches?,''
\newblock in {\em Proc. Inform. Theory and Appl. Workshop}, Feb. 2015, pp.
  90--97.

\bibitem{Park2016}
S.~Park and R.~W. Heath,
\newblock ``Spatial channel covariance estimation for {mmWave} hybrid {MIMO}
  architecture,''
\newblock in {\em Proc. Asilomar Conf. Sig., Sys., and Comp.}, Nov. 2016, pp.
  1424--1428.

\bibitem{Alkh2014}
A.~Alkhateeb, O.~El Ayach, G.~Leus, and R.~W. Heath,
\newblock ``Channel estimation and hybrid precoding for millimeter wave
  cellular systems,''
\newblock {\em IEEE J. Sel. Topics Sig. Process.}, vol. 8, no. 5, pp. 831--846,
  Oct. 2014.

\bibitem{Gao2016lcom}
Z.~Gao, C.~Hu, L.~Dai, and Z.~Wang,
\newblock ``Channel estimation for millimeter-wave massive {MIMO} with hybrid
  precoding over frequency-selective fading channels,''
\newblock {\em IEEE Commun. Lett.}, vol. 20, no. 6, pp. 1259--1262, June 2016.

\bibitem{Hagh2016wsa}
S.~Haghighatshoar and G.~Caire,
\newblock ``Enhancing the estimation of {mm-Wave} large array channels by
  exploiting spatio-temporal correlation and sparse scattering,''
\newblock in {\em Proc. Int. ITG Workshop on Smart Antennas}, Mar. 2016, pp.
  1--7.

\bibitem{Alkh2015icassp}
A.~Alkhateeb, G.~Leusz, and R.~W. Heath,
\newblock ``Compressed sensing based multi-user millimeter wave systems: How
  many measurements are needed?,''
\newblock in {\em Proc. IEEE Int. Conf. Acoust., Speech, Sig., Process.}, Apr.
  2015, pp. 2909--2913.

\bibitem{Dong2016wcsp}
Y.~Dong, C.~Chen, and Y.~Jin,
\newblock ``{AoAs} and {AoDs} estimation for sparse millimeter wave channels
  with one-bit {ADCs},''
\newblock in {\em Proc. Int. Conf. on Wireless Commun. Signal Process.}, Oct.
  2016, pp. 1--5.

\bibitem{Mo2016}
J.~Mo, P.~Schniter, and R.~W. Heath,
\newblock ``Channel estimation in broadband millimeter wave {MIMO} systems with
  few-bit {ADCs},''
\newblock {\em arXiv:1610.02735}, 2016.

\bibitem{Rusu2015}
C.~Rusu, R.~M{\'{e}}ndez-Rial, N.~Gonz{\'{a}}lez-Prelcic, and R.~W. Heath,
\newblock ``Adaptive one-bit compressive sensing with application to
  low-precision receivers at {mmWave},''
\newblock in {\em Proc. IEEE Global Commun. Conf.}, Dec. 2015, pp. 1--6.

\bibitem{Choi2016}
J.~Choi, J.~Mo, and R.~W. Heath,
\newblock ``Near maximum-likelihood detector and channel estimator for uplink
  multiuser massive {MIMO} systems with one-bit {ADCs},''
\newblock {\em IEEE Trans. Commun.}, vol. 64, no. 5, pp. 2005--2018, May 2016.

\bibitem{Rodriguez2016}
J.~Rodr{\'{i}}guez-Fern{\'{a}}ndez, N.~Gonz{\'{a}}lez-Prelcic, and R.~W. Heath,
\newblock ``Channel estimation in mixed hybrid-low resolution {MIMO}
  architectures for {mmWave} communication,''
\newblock in {\em Proc. Asilomar Conf. Sig., Sys., and Comp.}, Nov. 2016, pp.
  768--773.

\bibitem{sung2018iccasp}
J.~Sung, J.~Choi, and B.~L. Evans,
\newblock ``Channel estimation for hybrid beamforming millimeter wave
  communication systems with one-bit quantization,''
\newblock in {\em Proc. IEEE Int. Conf. Acoust., Speech, Sig., Process.},
\newblock
  \url{http://users.ece.utexas.edu/~bevans/papers/2018/channelEstOneBit/}, to
  be submitted.

\bibitem{Mo2014}
J.~Mo, P.~Schniter, N.~Gonz{\'{a}}lez-Prelcic, and R.~W. Heath,
\newblock ``Channel estimation in millimeter wave {MIMO} systems with one-bit
  quantization,''
\newblock in {\em Proc. Asilomar Conf. Sig., Sys., and Comp.}, Nov. 2014, pp.
  957--961.

\bibitem{rangan2014isit}
S.~Rangan, P.~Schniter, and A.~Fletcher,
\newblock ``On the convergence of approximate message passing with arbitrary
  matrices,''
\newblock in {\em Proc. IEEE Int. Symp. on Inform. Theory}, June 2014, pp.
  236--240.

\bibitem{projectcode}
J.~Sung and B.~L. Evans,
\newblock ``Wideband channel estimation for hybrid beamforming millimeter wave
  communication systems with low-resolution {ADCs},'' Software Release, Oct.
  13, 2017,
  \url{http://users.ece.utexas.edu/~bevans/projects/mimo/software/channel/}.

\bibitem{mcdonnell2008stochastic}
M.~D. McDonnell, N.~G. Stocks, C.~E.~M. Pearce, and D.~Abbott,
\newblock {\em Stochastic Resonance: From Suprathreshold Stochastic Resonance
  to Stochastic Signal Quantization},
\newblock Cambridge University Press, 2008.

\bibitem{Stockle2016}
C.~St{\"o}ckle, J.~Munir, A.~Mezghani, and J.~A. Nossek,
\newblock ``Channel estimation in massive {MIMO} systems using 1-bit
  quantization,''
\newblock in {\em Proc. IEEE Int. Work. Sig. Process. Adv. Wireless Commun.},
  July 2016, pp. 1--6.

\end{thebibliography}
\end{document}